\documentclass[aps, twocolumn, a4paper, superscriptaddress]{revtex4}
\usepackage{graphicx}
\usepackage{float}
\usepackage{color}
\usepackage{amsmath, amssymb}
\usepackage{lmodern}
\usepackage[T1]{fontenc}
\usepackage[utf8]{inputenc}
\usepackage[english]{babel}
 
\usepackage[urlcolor=blue,colorlinks,breaklinks=true]{hyperref}

\PassOptionsToPackage{obeyspaces,spaces,hyphens}{url}
 
 \usepackage{amsmath}

\begin{document}
 	\def\half{{1\over2}}
 	\def\shalf{\textstyle{{1\over2}}}
 	
 	\newcommand\lsim{\mathrel{\rlap{\lower4pt\hbox{\hskip1pt$\sim$}}
 			\raise1pt\hbox{$<$}}}
 	\newcommand\gsim{\mathrel{\rlap{\lower4pt\hbox{\hskip1pt$\sim$}}
 			\raise1pt\hbox{$>$}}}

\newcommand{\be}{\begin{equation}}
\newcommand{\ee}{\end{equation}}
\newcommand{\bq}{\begin{eqnarray}}
\newcommand{\eq}{\end{eqnarray}}
 	
\title{Domain walls and other defects in Eddington-inspired Born-Infeld gravity}
 	 	
\author{P.P. Avelino}
\email[Electronic address: ]{pedro.avelino@astro.up.pt}
\affiliation{Departamento de F\'{\i}sica e Astronomia, Faculdade de Ci\^encias, Universidade do Porto, Rua do Campo Alegre s/n, 4169-007 Porto, Portugal}
\affiliation{Instituto de Astrof\'{\i}sica e Ci\^encias do Espa{\c c}o, Universidade do Porto, CAUP, Rua das Estrelas, 4150-762 Porto, Portugal}
\affiliation{School of Physics and Astronomy, University of Birmingham,Birmingham, B15 2TT, United Kingdom}

\author{L. Sousa}
\email[Electronic address: ]{lara.sousa@astro.up.pt}
\affiliation{Instituto de Astrof\'{\i}sica e Ci\^encias do Espa{\c c}o, Universidade do Porto, CAUP, Rua das Estrelas, 4150-762 Porto, Portugal}
\date{\today}
\begin{abstract}

We investigate domain wall and other defect solutions in the weak-field limit of Eddington-inspired Born-Infeld gravity as a function of $\kappa$, the only additional parameter of the theory with respect to General Relativity. We determine, both analytically and numerically, the internal structure of domain walls, quantifying its dependency on $\kappa$ as well as the impact of such dependency on the value of the tension measured by an outside observer. We find that the pressure in the direction perpendicular to the domain wall can be, in contrast to the weak-field limit of General Relativity, significantly greater or smaller than zero, depending, respectively, on whether $\kappa$ is positive or negative. We further show that the generalized von Laue condition, which states that the average value of the perpendicular pressure is approximately equal to zero in the weak-field limit of General Relativity, does not generally hold in EiBI gravity not only for domain walls, but also in the case cosmic strings and spherically symmetric particles. We argue that a violation of the generalized von Laue condition should in general be expected in any theory of gravity whenever geometry plays a significant role in determining the defect structure.

\end{abstract}

\maketitle
 	
\section{Introduction}
\label{sec:intr}

Eddington-inspired Born-Infeld (EiBI) gravity \cite{Banados:2010ix} (see also  \cite{Deser:1998rj, Vollick:2003qp,Vollick:2005gc,Vollick:2006qd}) --- a particular example of a wider class of Ricci-based metric-affine theories of gravity \cite{Olmo:2011uz,Latorre:2017uve,Delhom:2019wir} --- has its roots on Born-Infeld non-linear electrodynamics \cite{Born:1934gh} (see \cite{BeltranJimenez:2017doy} for a recent review of Born–Infeld-inspired modifications of gravity). The equations of motion for the gravitational field are akin to the Einstein field equations, except for the replacement of the physical metric and the physical energy-momentum tensor by the so-called apparent ones \cite{Delsate:2012ky}. Although the apparent metric is, in the weak-field limit considered in the present paper, approximately equal to the Minkowski metric, the physical and apparent energy-momentum tensors may differ significantly inside matter \cite{Pani:2011mg,Casanellas:2011kf,Avelino:2012ge,Pani:2012qb,Avelino:2012qe,Pani:2012qd,Harko:2013wka}. This may result in a number of interesting effects, especially on microscopic scales where gravitational effects are often neglected. In this context, some cosmological and astrophysical singularities predicted to arise in the context of General Relativity could potentially be avoided \cite{Banados:2010ix,Avelino:2012ue,Scargill:2012kg,Bouhmadi-Lopez:2013lha,Bouhmadi-Lopez:2014jfa}.

EiBI gravity has a single additional parameter $\kappa$ with respect to General Relativity, thus allowing --- in combination with the other fundamental constants of the  theory (the speed of light in vacuum $c$ and Newton's gravitational constant $G$) --- for the definition of a fundamental length, time, mass and energy density, roughly determining the characteristic physical scales where significant deviations from General Relativity are expected to be found \cite{Avelino:2012ge}. Unless $|\kappa|$ is extremely close to zero, strong deviations from General Relativity are generally expected when considering extremely high energy densities such as those of the early universe \cite{Banados:2010ix,Avelino:2012ue,Scargill:2012kg}, the interior of compact astrophysical objects such as black holes \cite{Olmo:2013gqa,Sotani:2014lua, Wei:2014dka, Jana:2015cha, Avelino:2015fve,Avelino:2016kkj} and neutron stars \cite{Pani:2011mg,Pani:2012qb,Harko:2013wka}, or even inside subatomic particles such as the proton \cite{Avelino:2012qe}. 

In this paper, we consider the potential impact of EiBI gravity on the microscopic structure of topological defects, with a particular focus on domain walls. A key aspect that will be investigated in the present paper is the breakdown of the generalized von Laue condition, which states that the average value of the physical pressure in the direction perpendicular to the defect is approximately equal to zero in the weak-field limit of General Relativity (see \cite{doi:10.1002/andp.19113400808} for the original paper by von Laue, published in $1911$). This condition has been shown not only to apply to particles of fixed mass and structure, such as the proton \cite{Polyakov:2018zvc,Avelino:2019esh} (see \cite{Burkert:2018bqq} for a recent experimental test), but also to defects of co-dimension $D<N$ in $N+1$-dimensional space-times \cite{Avelino:2018qgt}. In this paper, we focus on domain wall solutions in three spatial dimensions ($N=3$, $D=1$) but we shall also discuss the case of cosmic strings ($N=3$, $D=2$) and spherically symmetric particles ($N=3$, $D=3$).

The outline of this paper is as follows. In Sec. \ref{sec:dw} we start by reviewing standard domain wall solutions in Minkowski space. In Sec. \ref{sec:eibi} we briefly describe the EiBI theory of gravity, giving particular relevance to the correspondence to General Relativity in terms of apparent tensor fields. In Sec. \ref{sec:dweibi} we study, both analytically and numerically, static domain wall solutions in the weak-field limit of EiBI gravity up to first order in $\kappa \rho$, giving particular emphasis to the breakdown of the generalized von Laue condition. In Sec. \ref{sec:oeibi} we extend the analysis of the previous section to cosmic strings, as well as spherically symmetric particles. We then conclude in Sec.  \ref{sec:conc}.

Here, we adopt a metric signature $(-,+,+,+)$. Greek indices and Latin indices $i$ and $j$ take the values $0,...,3$ and $1,...,3$, respectively. The Einstein summation convention will be used when a Greek index appears twice in a single term, once in an upper position (as superscript) and once in a lower position (as subscript). Also, we shall use units in which $c=8\pi G =1$.

\section{Domain walls in Minkowski space}
\label{sec:dw}

Consider a class of scalar field models described by the action
\be
S= \int d^4 x \sqrt {-g} \mathcal L\,,
\ee
where the Lagrangian $\mathcal L$ is given by
\be
\mathcal L = -\frac12 \partial_ \mu \phi  \partial^\mu \phi - V(\phi)\,.
\ee
Here, $\phi$ is a scalar field, $V(\phi)$ is the scalar field potential and $\partial_\mu$ represents a derivative with respect to the spacetime coordinate $x^\mu$. The components of the corresponding energy-momentum tensor are given by
\be
{T^\mu}_\nu=\partial^\mu  \phi \partial_\nu \phi +{\delta^\mu}_\nu \mathcal L\,,
\ee
where ${\delta^\mu}_\nu$ is the Kronecker delta.

In this paper, we shall consider scalar field models that admit domain wall solutions. For definiteness, we shall work with the potential
\be
V(\phi)=V_0 \left(\frac{\phi^2}{\eta^2} - 1 \right)^2\,, \label{potential}
\ee
where $V_0 = V(\phi=0)$ and $\phi=\pm \eta$ are its two degenerate minima. The model is, therefore,  invariant under transformations of the form $\phi \to -\phi$, thus possessing a $Z_2$ symmetry. 

If the gravitational field can be neglected within the domain wall, the metric is essentially Minkowskian. In this case, the equation of motion of a static planar domain wall oriented perpendicularly to the $z$ axis is simply given by
\be
\phi''=\frac{dV}{d\phi}\,, \label{phiev}
\ee
where a prime represents a derivative with respect to $z$.
This equation may be integrated to obtain
\be
\label{phiprime}
\frac{\phi'^2}{2}=V(\phi)\,.
\ee
Here, we have taken the integration constant to be equal to zero, since $\phi' \to 0$ and $V(\phi) \to 0$ as $\phi \to \pm \eta$ in the case of a static planar domain wall. 

The following static planar domain wall solution,
\be
\label{phidw}
\phi=\eta \tanh \left(\frac{z}{\delta}\right)\,,
\ee
with $\delta=2^{-1/2} \eta V_0^{-1/2}$, may be easily computed by imposing that $\phi=0$ at $z=0$. 

In a Minkowski spacetime, the off-diagonal components of the energy-momentum tensor vanish, while the diagonal components are equal to
\bq
{T^0}_0&=&{T^x}_x={T^y}_y=-\rho\,,\label{T1}\\
{T^z}_z&=&p_z\label{T2}\,,
\eq
with $\rho=\phi'^2/2+V(\phi)$ and $p_z=\phi'^2/2-V(\phi)=0$. Hence, given Eqs.~\eqref{potential} and ~\eqref{phiprime}-\eqref{phidw}, one may find an analytical expression for the energy density of a static planar domain wall:
\be
\rho(|z|)=2 V_0 \, {\rm sech}^4 \left(\frac{z}{\delta}\right).\label{rhozdw0}
\ee

\section{EiBI gravity  \label{sec:eibi}}

The EiBI gravity action is given by
\be
S=\frac{2}{\kappa}\int d^{4}x\left[\sqrt{\left|{\rm det}(g_{\mu\nu}+\kappa R_{\mu\nu})\right|}-\lambda\sqrt{|g|}\right]+S_M\,,\label{eq:EddingtonBornInfeld Action}
\ee
where $g_{\mu\nu}$ are the components of the physical metric, $g \equiv {\rm det}(g_{\mu\nu})$ is the determinant of $g_{\mu\nu}$, $R_{\mu\nu}$ is the symmetric Ricci tensor built from the connection  $\Gamma$ (see \cite{BeltranJimenez:2019acz} for a discussion of why only the symmetric part should be considered), $S_M$ is the standard action associated with the matter fields, and $\kappa$ is the only additional parameter of the theory with respect to General Relativity. Without loss of generality, we set $\lambda=1$ (since the changes associated to a different value of $\lambda$ can be incorporated into the energy-momentum tensor).

We shall consider the Palatini formulation of EiBI gravity in which the metric and the connection are treated as independent fields. The equations of motion
\bq
q_{\mu\nu}&=&g_{\mu\nu}+\kappa R_{\mu\nu}\,,\label{eq:ConnectionEquationOfMotion}\\
\sqrt{|q|}q^{\mu\nu}&=&\sqrt{|g|}g^{\mu\nu}-\kappa\sqrt{|g|}T^{\mu\nu}\,,\label{eq:MetricEquationOfMotion}
\eq
can then be obtained by considering the variation of the action with respect to the connection and the physical metric, respectively. Here, $T^{\mu \nu}$ are the components of the energy-momentum tensor, $q_{\mu\nu}$ is an auxiliary (apparent) metric related to the original connection by 
\be
\Gamma^{\gamma}_{\mu\nu} = \frac{1}{2} q^{\gamma\zeta}( \partial_\nu q_{\zeta\mu} + \partial_\mu q_{\zeta\nu}- \partial_\zeta q_{\mu\nu})\,, \label{connection}
\ee
$q^{\mu\nu}$ is the inverse of $q_{\mu\nu}$, and $q={\rm det}(q_{\mu\nu})$. 

Eqs.~(\ref{eq:ConnectionEquationOfMotion}) and (\ref{eq:MetricEquationOfMotion}) may be combined to obtain the following second-order field equations
\be
{{\mathcal G}^\mu}_\nu \equiv {{{\mathcal R}}^\mu}_\nu -\frac12 {\mathcal R} {\delta^\mu}_\nu  ={{\mathcal T}^{\mu}}_{\nu}\,,\label{eq:EquationOfMotionComb1}
\ee
with 
\bq
{{{\mathcal R}}^\mu}_\nu &\equiv& q^{\mu \zeta} R_{\zeta \nu} ={\Theta^{\mu}}_{\nu}\,,\label{eq:EquationOfMotionComb}\\
{{\mathcal T}^{\mu}}_{\nu} &\equiv& {\Theta^{\mu}}_{\nu}-\frac12\Theta {\delta^\mu}_\nu\,,\label{eq:aThetamunu}\\
{\Theta^{\mu}}_{\nu}&\equiv&\frac{1}{{\kappa}}\left(1-\tau\right){\delta^\mu}_\nu+\tau{T^{\mu}}_{\nu} \,,\label{eq:Thetamunu}\\
\Theta &\equiv&  {\Theta^{\mu}}_{\mu}\,\label{eq:Theta}\\
\tau&\equiv&\sqrt{{\frac{g}{q}}}= \left[ \det( {\delta^\mu}_\nu -  \kappa {T^\mu}_\nu ) \right]^{-\frac{1}{2}} \label{eq:tau}\,.
\eq
Here, ${{\mathcal G}^\mu}_\nu$ and ${{\mathcal T}^{\mu}}_{\nu}$ are the components of the so-called apparent Einstein tensor and of the apparent energy-momentum tensor, respectively. The equations of motion of EiBI gravity are analogous to those of General Relativity, except for the fact that the apparent Einstein tensor is calculated from the apparent metric --- instead of the physical metric --- and that the physical energy-momentum tensor is replaced by the apparent one.

The weak-field limit of EiBI gravity is defined as the limit where the apparent metric is approximately Minkowskian. Note, however, that inside matter the physical and apparent metrics do not coincide, even in this limit: as a matter of fact, the physical metric may be significantly different from the Minkowski metric if the energy density is large. As we will show in the following sections, this may have an impact on the structure of topological defects, such as domain walls or cosmic strings.

\section{Domain walls in EiBI gravity}
\label{sec:dweibi}

Given the non-zero components of the proper energy-momentum tensor of a domain wall, defined in Eqs. \eqref{T1} and \eqref{T2}, $\tau$ may be written as
\be
\tau = \left[(1+\kappa \rho)^3 (1-\kappa p_z)\right]^{-1/2}\,.
\label{taudef}
\ee

The non-zero components of the apparent energy-momentum tensor may then be calculated using Eqs. \eqref{eq:aThetamunu}, \eqref{eq:Thetamunu} and \eqref{eq:Theta}. They are given by
\bq
{{\mathcal T}^{0}}_{0} &=& {{\mathcal T}^{x}}_{x}=  {{\mathcal T}^{y}}_{y}=-\frac{1}{{\kappa}}\left(1-\tau\right) + \frac12 \tau (\rho - p_z)\,, \label{appT00}\\
{{\mathcal T}^{z}}_{z} &=& -\frac{1}{{\kappa}}\left(1-\tau\right) + \frac12 \tau (3\rho + p_z)\,.\label{appTzz}
\eq

Here, we shall work in the weak-field limit of EiBI gravity, neglecting Newtonian and post-Newtonian corrections. In this limit, the apparent metric inside the wall is well approximated by the Minkowski metric and the conservation of the apparent energy-momentum tensor implies that ${{\mathcal T}^{z}}_{z,z} = 0$. Hence, the apparent pressure should be such that
\be
p_{z\rm [A]} \equiv {{\mathcal T}^{z}}_{z} = 0 \label{pzstar}
\ee 
everywhere, if one assumes that it vanishes at an infinite distance from the domain wall. It is simple to show, using Eqs. \eqref{appTzz} and \eqref{pzstar}, that the domain wall solution then satisfies
\be
\tau=\frac{1}{1+\kappa \left(3 \rho + p_z\right)/2}\,. \label{taunew}
\ee
On the other hand, by substituting Eq.~\eqref{taunew} into Eq.~\eqref{appT00}, and using Eqs. \eqref{appTzz} and \eqref{pzstar}, one obtains an expression for the apparent energy density $\rho_{\rm [A]}$ in terms of the physical energy density $\rho$ and of the perpendicular pressure $p_z$:
\bq
\rho_{\rm [A]} &\equiv& - {{\mathcal T}^{0}}_{0} = -p_{z\rm [A]} + \tau \left( \rho + p_z \right)\nonumber \\
&=& \tau \left( \rho + p_z \right)=\frac{\rho + p_z}{1+\kappa \left(3 \rho + p_z\right)/2}\,.
\eq

Expanding Eq.~\eqref{taudef} up to second order in $\kappa {T^\mu}_\nu$, one finds that
\bq
\tau &=& 1-\frac32 \kappa \rho + \frac12 \kappa p_z +\frac{15}{8} \kappa^2 \rho^2  - \frac34 \kappa^2 \rho p_z \nonumber  \\
&+& \frac38 \kappa^2 p_z^2\,. \label{tauapprox}
\eq
One may show, using Eqs.~\eqref{appTzz}, \eqref{pzstar} and \eqref{tauapprox},  that
\be
0 = p_{\rm [A]} \equiv {{\mathcal T}^{z}}_{z} =
p_z -\frac{3}{8} \kappa \rho^2 \,, \label{tzzapprox}
\ee
up to first order in $\kappa \rho$ and so $p_z$ and $p_{z\rm [A]}$ coincide only up to zeroth order in $\kappa \rho$. Notice that the second term in Eq. \eqref{tzzapprox} represents an effective gravitational pressure in the direction perpendicular to the domain wall that is symmetric to the physical pressure $p_z$. Hence, an approximate analytical solution for $p_z(z)$, valid up to first order in $\kappa \rho$, may be obtained by substituting the $\kappa=0$ solution for $\rho$ given in Eq.~\eqref{rhozdw0} into Eq.~ \eqref{tzzapprox}: 
\be
p_z^{\rm analytical}(|z|)=\frac32 \kappa V_0^2 \, {\rm sech}^8 \left(\frac{z}{\delta}\right)\,.\label{papprox}
\ee
On the other hand, using Eqs.~\eqref{appT00}, \eqref{tauapprox} and~\eqref{tzzapprox}, one may show that the relation between the physical and apparent energy densities is approximately given by
\be
\rho_{\rm [A]} = \rho -\frac98 \kappa \rho^2 \,, \label{rhoA}
\ee
up to first order in $\kappa \rho$.

Differentiating Eq. \eqref{appTzz} with respect to $z$, one finds
\be
\frac{\tau'}{\tau}\left[\frac{1}{\kappa} + \frac12 \left( 3 \rho+p_z\right) \right]+ \frac12 \left( 3 \rho'+p_z'\right) = 0\,. \label{d23dz}
\ee
On the other hand, by differentiating Eq. \eqref{taudef} with respect to $z$ one obtains
\be
(\ln \tau)'=\frac{\tau'}{\tau}=-\frac{3}{2} \frac{\kappa \rho'}{1+\kappa \rho}  +\frac{1}{2} \frac{\kappa p_z '}{1-\kappa p_z}\,.
\ee
One may then show that
\be
p_z'\left[1+f(\rho,p_z)\right]=\rho' h(\rho,p_z)\,, \label{pzprime}
\ee
where
\bq
f(\rho,p_z)&=&\frac14\left[\kappa \left(7\rho-p_z\right)+\kappa^2 \rho (3 \rho-p_z)\right]  \,, \label{fdef}\\
h(\rho,p_z)&=&\frac34 \kappa \left(\rho + p_z\right)\left(1-\kappa p_z\right)\label{gdef}\,,
\eq
and
\be
\rho= g^{zz} \frac{\phi'^2}{2} + V(\phi)  \,, \qquad p_z= g^{zz} \frac{\phi'^2}{2} - V(\phi)\,. \label{rhopz}
\ee

Equations \eqref{eq:MetricEquationOfMotion} and \eqref{taunew},  imply that
\be
g^{zz}=q^{zz} \tau^{-1} + \kappa T^{zz} = 1 + \frac32 \kappa \rho \ee
up to first order in $\kappa \rho$ (where we have used the fact that $T^{zz} = p_z = 0$ up to zeroth order in $\kappa \rho$). Using Eq. \eqref{rhopz}, one finds that
\be 
g^{zz} =  1 + \frac32 \kappa  \left[\frac12 \phi'^2 + V(\phi)\right]\,,\label{gzz}
\ee
up to first order in $\kappa \rho$. By substituting  Eqs. \eqref{rhopz} and \eqref{gzz} into Eq. \eqref{pzprime}, one obtains the following second order equation for $\phi$
\be
\phi''=\beta (\phi,\phi') \frac{dV}{d \phi} \,, \label{phiev1}
\ee
with
\be
\beta  = \frac{\left(1+f\right)\left(1-3\kappa \phi'^2/4\right)+h\left(1+3\kappa \phi'^2/4\right)}{\left(1+f-h\right)\left(1+3\kappa \left(\phi'^2+V\right)/2\right)}\,,
\ee
Here $f(\phi,\phi')$ and $h(\phi,\phi')$ may be computed using Eqs. \eqref{fdef}, \eqref{gdef}, \eqref{rhopz} and \eqref{gzz}. One has $\beta=1$ up to zeroth order in $\kappa \rho$ and thus this equation reduces to Eq. \eqref{phiev} in the $\kappa \rho \to 0$ limit. However, away from it, these two equations may lead to significantly different domain wall solutions.

Notice that Eq. \eqref{rhopz} implies that $\rho-p_z=2 V_0$ at the domain wall core (defined by $\phi=0$). Equation \eqref{tzzapprox} then implies that the energy density at the domain wall core is equal to
\be
\label{arho0}
\rho_0^{\rm analytical} = 2 V_0 \left(1 + \frac34 \kappa V_0\right)\,, \ee
up to first order in $\kappa V_0$. Similarly, using Eq. \eqref{rhoA}, one may now find an analogous approximation to the apparent energy density at the core of the domain wall:
\be
\rho^{\rm analytical}_{0\rm [A]} = 2 V_0 \left(1 - \frac32 \kappa V_0\right)\,. \label{arho0A}
\ee
Using Eqs. \eqref{rhopz}, \eqref{gzz} and \eqref{arho0}, and taking into account that $\phi'^2/2=V$ up to order zero in $\kappa \rho$, one finds that 
\be
\phi'_0  \equiv \phi'(z=0) = \sqrt{2 V_0 \left(1 - 3 \kappa V_0/2\right)}\,, \label{aphi0}
\ee
and as a result we should have, in the vicinity of the domain wall core, that
\be
\phi = \sqrt{2 V_0 \left(1 - 3 \kappa V_0/2\right)} \, z\,, \label{aphi01}
\ee
where we have used the fact that $\phi(z=0)=0$ (again, these approximations are valid up to first order in $\kappa \rho$). Hence, we may conclude that, for $\kappa \rho \ll 1$, the slope of the $\phi$ profile close to the domain wall core is a decreasing function of $\kappa$.

\begin{figure} 
	         \begin{minipage}{1.\linewidth}  

               \rotatebox{0}{\includegraphics[width=0.96\linewidth]{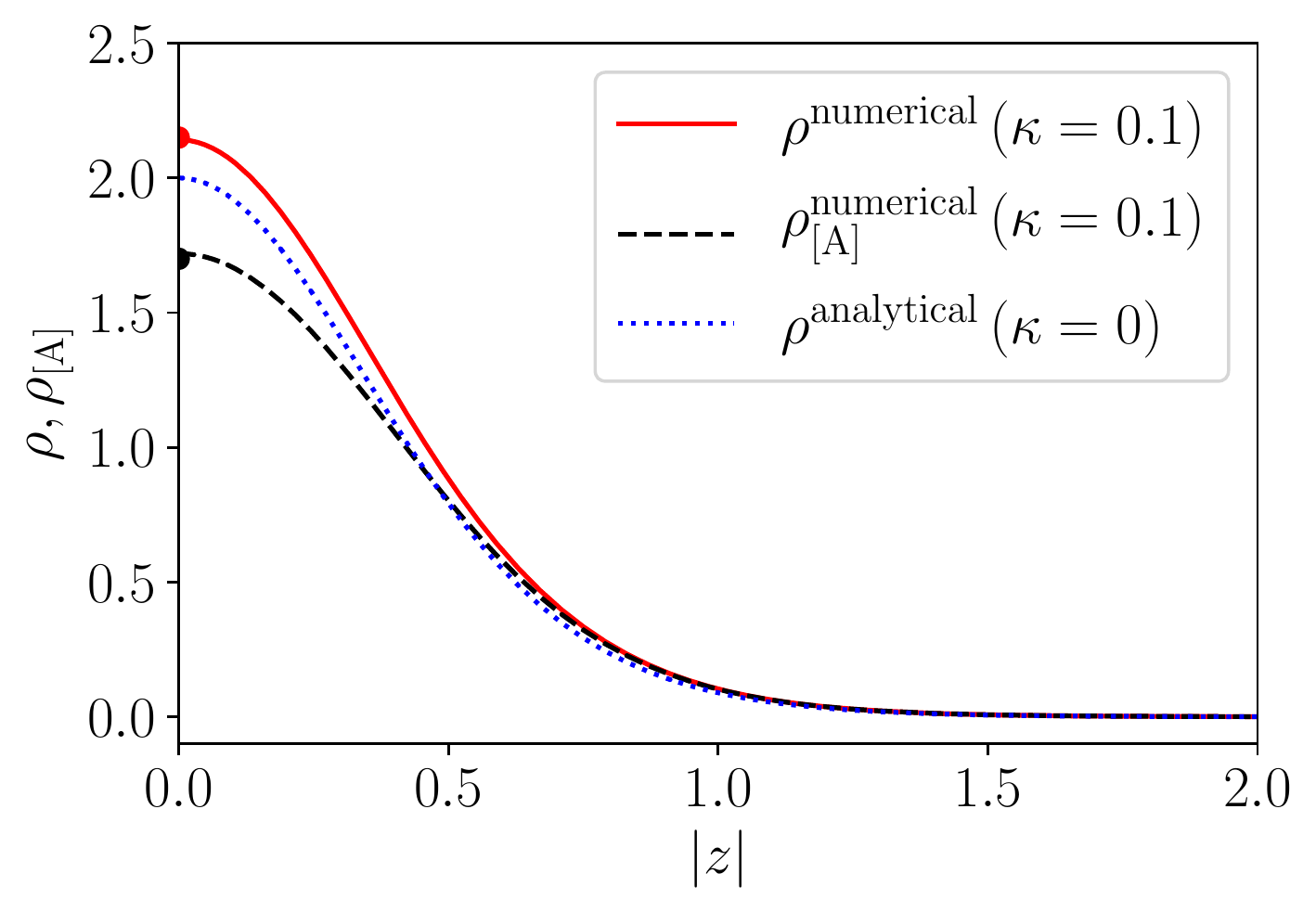}}
               \rotatebox{0}{\includegraphics[width=0.96\linewidth]{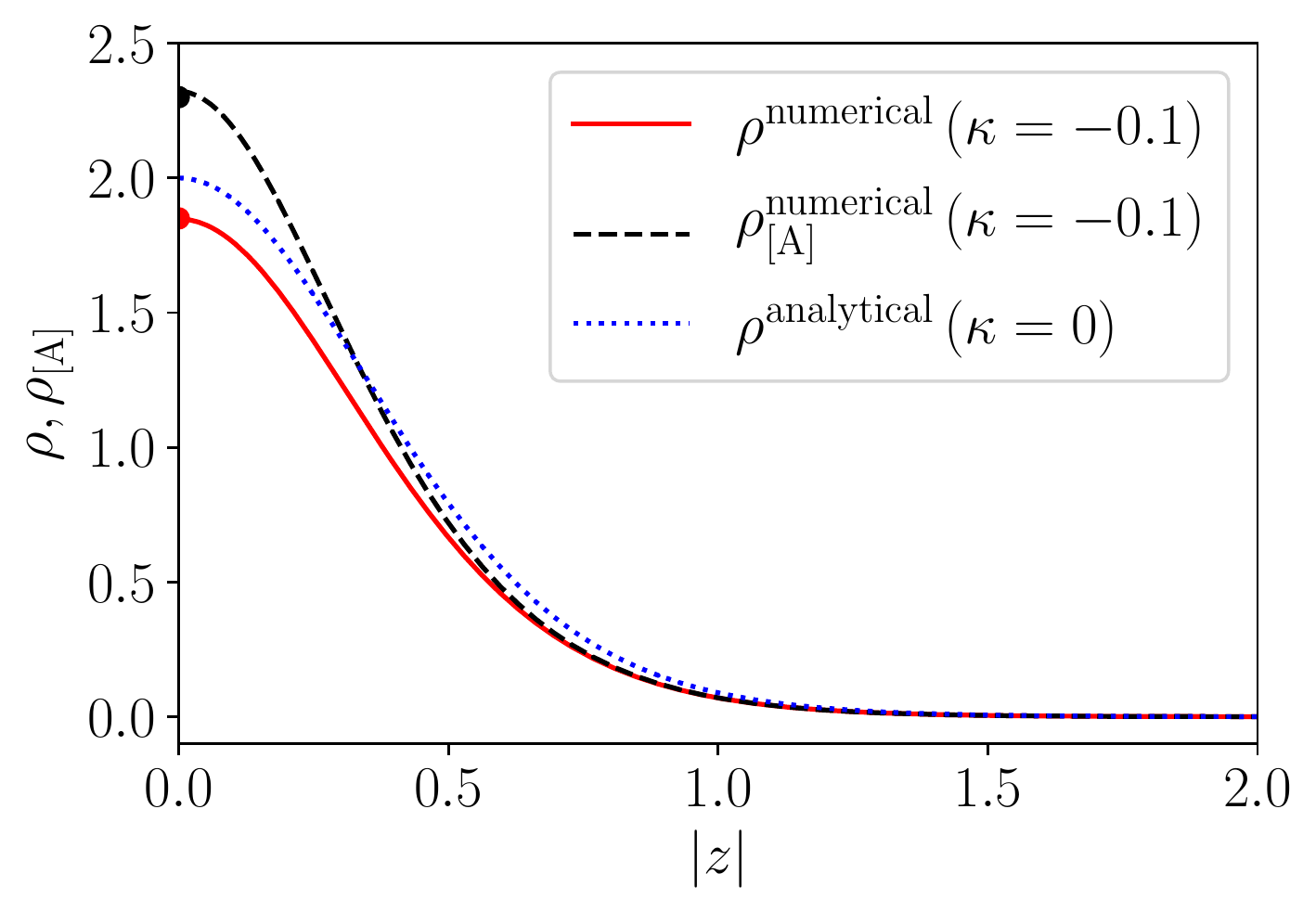}}
         \end{minipage}
	\caption{Numerical approximations to the physical energy density $\rho$ (red solid line) and the apparent energy density $\rho_{\rm [A]}$ (black dashed line) as a function of the distance $|z|$ to the domain wall core, for $\kappa=0.1$ (top panel) and $\kappa=-0.1$ (bottom panel). The red and black circles at $z=0$ represent the analytical estimates of the core physical and apparent energy densities of the domain wall given by Eqs.~\eqref{arho0} and \eqref{arho0A}, respectively. The analytical solution for the physical energy density for the $\kappa=0$ case (blue dotted line) is also shown for comparison [see Eq. \eqref{papprox}].}
	\label{fig1}
\end{figure}

\begin{figure} 
	         \begin{minipage}{1.\linewidth}  

               \rotatebox{0}{\includegraphics[width=1\linewidth]{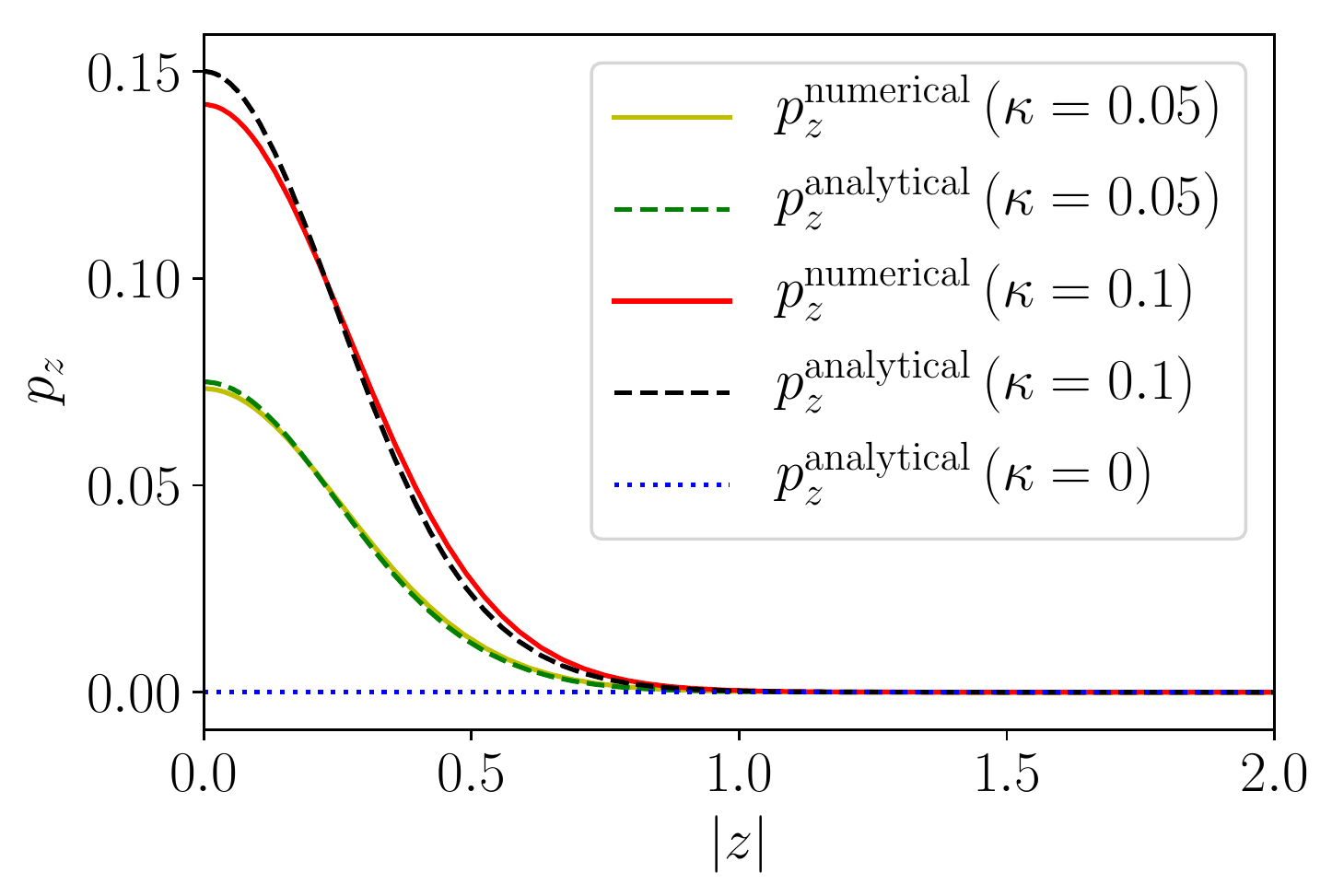}}
               \rotatebox{0}{\includegraphics[width=1\linewidth]{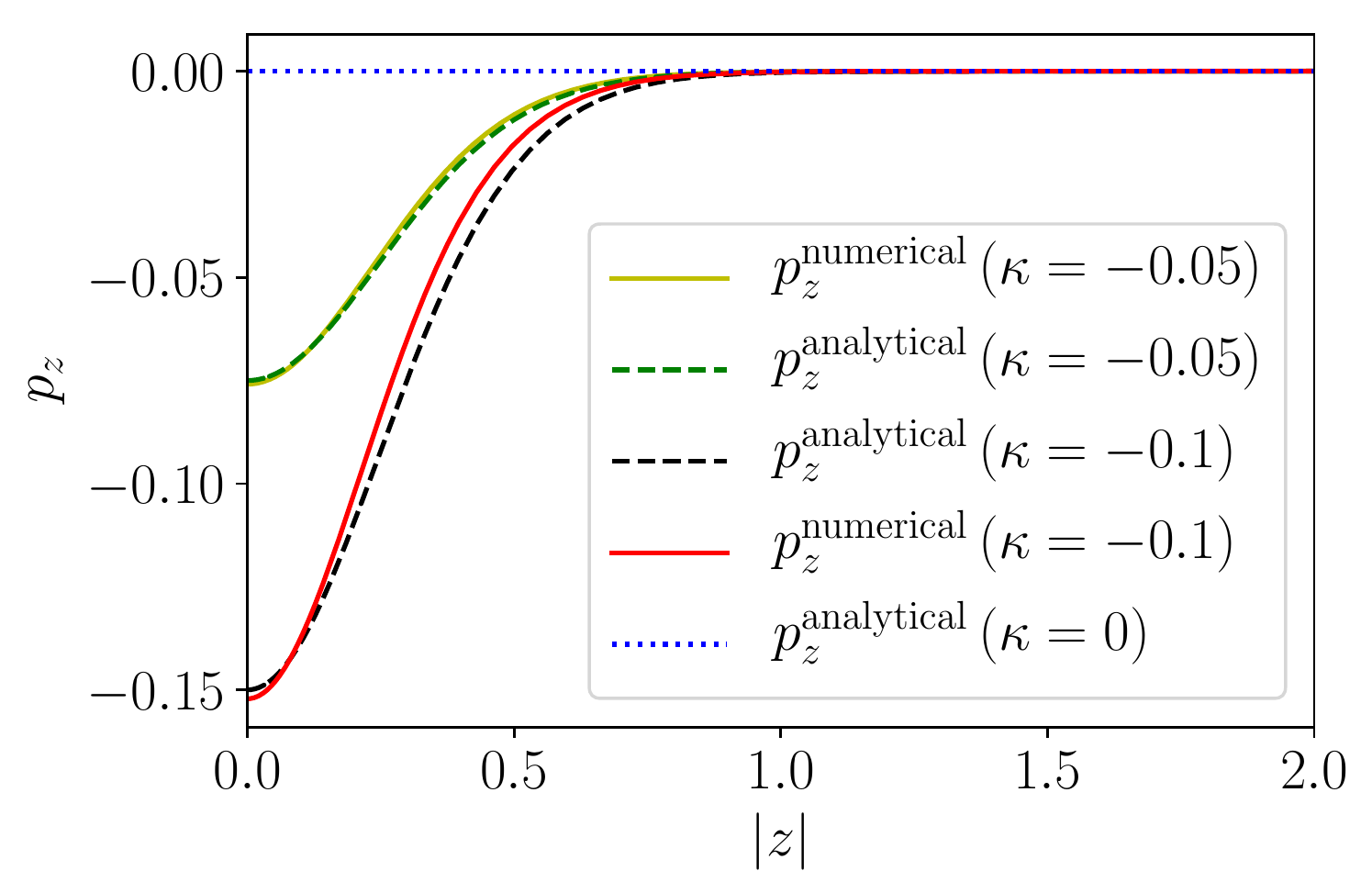}}
         \end{minipage}
	\caption{Analytical (dashed lines) and numerical (solid lines) approximations to the physical pressure $p_z$ as a function of the distance $|z|$ to the domain wall core, for $\kappa=0.05, 0.1$ (top panel) and $\kappa=-0.05,-0.1$ (bottom panel). The value of the pressure in the $\kappa=0$ case is given by the blue dotted line. Notice that, for $\kappa \rho \neq 0$, the pressure near the domain wall core deviates from zero, and that the analytical and numerical approximations provide comparable results both for $|\kappa|=0.05$ and $|\kappa|=0.1$.}
	\label{fig2}
\end{figure}

These analytical approximations --- unlike the approximation for the apparent pressure in Eq.~\eqref{papprox} --- are only valid at the domain wall core. However, one may solve  Eq.~\eqref{phiev1}
to obtain numerical solutions for $\rho(|z|)$ and $\rho_{\rm [A]}(|z|)$ that are valid up to first order in $\kappa \rho$. Here, we shall use units in which $V_0=1$ and we shall again assume that the domain wall core is located at $z=0$ (or, equivalently, that $\phi=0$ at $z=0$) and that $\eta=1$. Our results are independent of this choice, except for the fact that $|z|$ in Figs. \ref{fig1} and \ref{fig2} should be replaced by $|z|/\eta$ for other choices of $\eta$ . The value of $\phi'_0$ is found by numerically evaluating the root of the equation ${{\mathcal T}^{z}}_{z}=0$. 

\begin{figure} 
	         \begin{minipage}{1.\linewidth}  

               \rotatebox{0}{\includegraphics[width=1\linewidth]{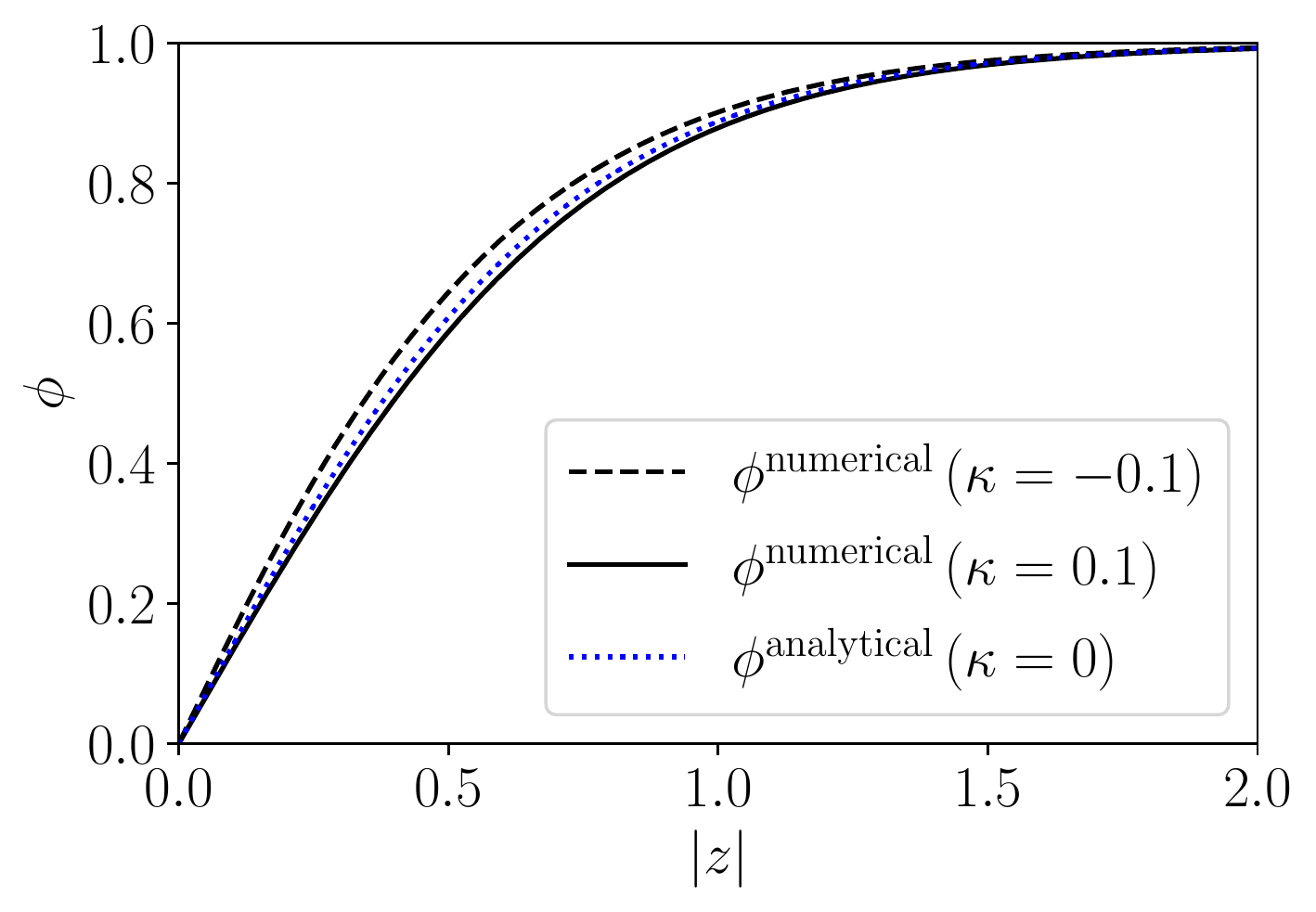}}
         \end{minipage}
	\caption{Numerical solution for the scalar field $\phi$ as a function of the distance $|z|$ to the domain wall core, for $\kappa=-0.1$ and $\kappa=0.1$ (black dashed and solid lines, respectively). The analytical solution with $\kappa=0$ is also shown for comparison (blue dotted line).}
	\label{fig3}
\end{figure}

Figure \ref{fig1} shows the numerical approximations to the physical and apparent energy densities (red solid and black dashed lines, respectively) for $\kappa=0.1$ (top panel) and $\kappa=-0.1$ (bottom panel). The red and black filled circles at $z=0$ represent the analytical estimates of the physical and apparent energydensities at the domain wall core given by Eqs. \eqref{arho0} and \eqref{arho0A}, respectively. The analytical solution for the physical energy density in the $\kappa=0$ case (blue dotted line) is also shown for comparison. Figure \ref{fig1} shows that while the physical energy density $\rho(|z|)$ is an increasing function of $\kappa$ in the regime where $|\kappa| \rho \ll 1$, the same does not happen to the apparent energy density $\rho_{\rm [A]}(|z|)$ sufficiently near the domain core. In fact, we verified numerically that, up to first order in $\kappa \rho$, the apparent domain wall tension --- the tension measured by an observer outside the domain wall --- is approximately given by
\be
\sigma_{[A]} \equiv \int_{-\infty}^\infty \rho_{[A]}(z) dz = {\widehat \sigma}_{[A]}   (1- \alpha \kappa {\widehat \sigma}_{[A]}) \,,
\ee
with ${\widehat \sigma}_{[A]}=\sigma_{[A]}(\kappa=0)=2 \sqrt{2} V_0^{1/2} \eta/3=2 \sqrt{2}/3$ and $\alpha \sim 0.5$. Hence,  $\sigma_{[A]}$ is a decreasing function of $\kappa$ in the regime where $|\kappa| \rho \ll 1$, in agreement with the results displayed in Fig. \ref{fig1}. An observer outside domain wall thus ``sees'' what appears to be a standard domain wall with a smaller apparent tension
if gravity is attractive (implying $p_z>0$ and a negative effective gravitational pressure) than if it is repulsive (implying $p_z<0$ and a positive effective gravitational pressure). Figure \ref{fig1} also shows the good agreement between the analytical estimates for physical and apparent energy densities at the core of the domain wall (red and black filled circles at $z=0$, respectively) and the corresponding numerical results, for $\kappa=0.1$ and $\kappa=-0.1$.

Figure \ref{fig2} displays the analytical (dashed lines) and numerical (solid lines) approximations to the physical pressure $p_z$ as a function of the distance $|z|$ to the domain wall core, for $\kappa=0.05, 0.1$ (top panel) and $\kappa=-0.05,-0.1$ (bottom panel). The value of the pressure in the $\kappa=0$ case --- $p_z=0$ for all $z$ --- is given by the blue dotted line. As expected, $p_z^{\rm numerical}$ is always positive or always negative, depending respectively on whether $\kappa>0$ (top panel) or $\kappa<0$ (bottom panel), in agreement with analytical approximation given in Eq.  \eqref{papprox}. Figure \ref{fig2} also shows that the analytical and numerical approximations to the physical pressure $p_z$ nearly coincide for $|\kappa|=0.05$ --- the accuracy of the analytical and numerical approximations are both decreasing functions of $|\kappa|$, thus explaining the larger discrepancies obtained for $|\kappa|=0.1$. 


Figure \ref{fig3} shows the numerical solution for the scalar field $\phi$ as a function of the distance $|z|$ to the domain wall core, considering $\kappa=-0.1$ and $\kappa=0.1$ (black dashed and solid lines, respectively). The analytical solution with $\kappa=0$ --- given by Eq.~\eqref{phidw} --- is also shown for comparison (blue dotted line). Figure \ref{fig3} shows that for $\kappa \in [-0.1,0.1]$ the scalar field profile is not significantly affected. Notice, however, that the slope of the scalar field profile (equal to $\phi'$) is always larger for $\kappa < 0$ than for $\kappa > 0$. This is in agreement with the analytical approximation for the slope of the scalar field profile in the vicinity of the domain wall core given in Eqs. \eqref{aphi0} and \eqref{aphi01}.

\section{Other defects in EiBI gravity}
\label{sec:oeibi}

The results of the previous section show that EiBI gravity can have a significant impact on the microscopic structure of domain walls. It not only affects the energy density of the domain walls (and thus its tension) but also results in the wall acquiring an internal perpendicular pressure (which is negligible in the weak-field limit of General Relativity). In this section, we shall investigate whether one should expect similar effects for other topological defects.

\subsection{A Generalized von Laue condition}
Let us consider a topological defect of co-dimension $D\le N$ in a $N+1$-dimensional Minkowski background and let us assume that its core is defined by the condition $x^i=0$, where $x^i$, with $i=1,...,p$, are spatial cartesian coordinates and $p=N-D$ is the dimensionality of the defect. It was shown in~\cite{Avelino:2018qgt}, using a Derrick-type argument~\cite{Derrick:1964ww}, that the volume average of the pressure along the perpendicular dimensions for stable localized defects of co-dimension $D$ --- i. e., for static defects with a finite size in the $D$ co-dimensions --- vanishes. In other words, we should have that
\be 
\int T^{ll}d^Dx=0\,,\mbox{for all }l=p+1,\cdots,N\,.
\label{gen-vonL}
\ee
This condition --- which is equivalent to requiring that the defect solution is an extremum of the energy --- is not sufficient to ensure its stability: any solution in static equilibrium must minimize energy too. However it is a necessary condition which applies not only to models described by scalar field multiplets, but also to models that include higher order tensor fields and it is expected to be verified whenever the gravitational field has a negligible impact on the defect internal structure. It is thus a pretty generic requirement for a stable defect solution in General Relativity. For $D=N$, this condition is equivalent to the von Laue condition~\cite{doi:10.1002/andp.19113400808} --- whereby the average pressure inside composed particles is equal to zero --- but extends its scope to generic solitonic particles and defects. In this sense, Eq.~\eqref{gen-vonL} may be regarded as a generalization of the von Laue condition to defects of arbitrary dimensionality.

According to this generalized von Laue condition, the volume average of the perpendicular pressure of domain walls (which are defects with N=3 and D=1) should vanish. However, the results of Sec.~\ref{sec:dweibi} show that this condition is violated in the presence of EiBI gravity. As discussed previously, EiBI gravity is akin to General Relativity with an apparent metric, and an exotic source (described by the apparent energy-momentum tensor). As a result, the preservation of the generalized von Laue condition in the weak-field limit of EiBI gravity (where the apparent metric is essentially Minkowskian) would require it to be expressed in terms of the apparent pressure instead of the physical one. In this sense, one may say that the generalized von Laue condition is \textit{physically} violated but \textit{apparently} satisfied.

In the following, we shall extend some of the work of the previous section to other types of defects. In particular, we shall investigate whether the violation of the generalized von Laue condition, found in the previous section for domain walls, also happens in the case of cosmic strings and spherically symmetric particles.

\subsection{Cosmic strings}
Let us start by considering cosmic strings (characterized by $N=3$ and $D=2$) with localized energy which may, in principle, be treated effectively as $1+1$-dimensional objects on sufficiently large scales. For a static straight cosmic string oriented along the $z$ direction, one has~\cite{Vilenkin:2000jqa}:
\be
{T^0}_0={T^z}_z=-\rho\,,\quad {T^r}_r=p_r\,,\quad {T^\phi}_\phi=p_\phi\,,
\ee
where $p_r$ and and $p_\phi$ represent the pressure in the $r$ and $\phi$ directions, ($r$,$\phi$,$z$) are cylindrical spatial coordinates, and all the off-diagonal components vanish. We then have that
\be
\tau = \left[(1+\kappa \rho)^2 (1-\kappa p_r) (1-\kappa p_\phi) \right]^{-1/2}\,.
\ee
 
The components of the apparent energy-momentum tensor are such that

\bq
{{\mathcal T}^{0}}_{0} &=& {{\mathcal T}^{z}}_{z}=-\frac{1}{{\kappa}}\left(1-\tau\right) - \frac{1}{2}\tau (p_r +  p_\phi)\,,\\
{{\mathcal T}^{r}}_{r} &=& -\frac{1}{{\kappa}}\left(1-\tau\right) +  \frac{1}{2}\tau (2\rho + p_r-p_\phi)\,,\\
{{\mathcal T}^{\phi}}_{\phi} &=& -\frac{1}{{\kappa}}\left(1-\tau\right) + \frac{1}{2}  \tau (2\rho - p_r+p_\phi)\,,
\eq
and thus the apparent pressure in the perpendicular direction is given by
\bq
p_{\perp \rm [A]} &\equiv& \frac12\left( {{\mathcal T}^{r}}_{r} + {{\mathcal T}^{\phi}}_{\phi} \right)= 
 -\frac{1}{{\kappa}}\left(1-\tau\right) + \tau \rho \nonumber \\
& = & \frac{1}{\kappa}\left\{\left[\left(1-\kappa p_r\right)\left(1-\kappa p_\phi\right)\right]^{-1/2}-1\right\}\,.
\eq
 Notice that the apparent perpendicular pressure is independent of $\rho$ and, as a consequence, this does not result in an additional pressure of the form of that presented in Eq.~\eqref{tzzapprox}.
 
 As a matter of fact, one has that, up to first order in $\kappa {T^\mu}_\nu$, 
\be 
 p_{\perp \rm [A]} =  p_\perp +\frac{3}{2}\kappa p_\perp^2-\frac{1}{2}\kappa p_\perp p_\phi\,,
 \ee
where $p_\perp = (p_r+p_\phi)/2$, and therefore the apparent pressure will equal zero provided that $p_r$ and $p_\phi$ are also vanishing. Note however that this is not necessarily the case: for Abelian-Higgs strings, the radial and azimuthal pressures vanish in Minkowski space only at critical coupling --- i.e., when the masses of the vector particle and the Higgs boson are exactly equal~\cite{deVega:1976xbp,anderson2002mathematical}. If this is not the case, $p_r$ and $p_\phi$ only vanish when averaged over the string cross section and, therefore, corrections to the von Laue condition may arise.

Although there are no violations of the generalized von Laue condition in the case of cosmic strings at critical coupling, the structure of a string in EiBI gravity is, in any case, different from that of strings in General Relativity. Up to second order in $\kappa {T^\mu}_\nu$, we have that 
\bq
\tau &=& 1-\kappa \left(\rho - p_\perp\right) + \frac{\kappa^2}{2} \left(2 \rho^2 - 2 \rho p_\perp + 3 p_\perp^2 \right.  \nonumber  \\
&-& \left.  p_r p_\phi \right)\,,
\eq
and, as a result, the apparent energy density of the string is given by
\be 
 \rho_{\perp \rm [A]}=-{{\mathcal T}^0}_0 = \rho-\kappa \rho^2\,,
\ee 
up to first order in $\kappa \rho$. So the string would have an apparent energy density that differs from the physical one, being  smaller (larger) than the physical energy density for positive (negative) values of $\kappa$.

\subsection{Spherically symmetric particles}

Let us now consider the case of spherically symmetric particles, with $N=D=3$, described by the following energy-momentum tensor:
\be
{T^0}_0=-\rho\,,\quad {T^r}_r=p_r\,,\quad {T^\phi}_\phi={T^\theta}_\theta=p_\phi\,,
\ee
where $(r,\theta,\phi)$ are spherical coordinates, $p_r$ and $p_\phi$ represent, respectively, the pressure in the $r$ and $\phi$ directions, and the equality of ${T^\theta}_\theta$ and ${T^\phi}_\phi$ stems from the spherical symmetry (all other components of the energy-momentum tensor vanish). We then have that
\be
\tau = \left[(1+\kappa \rho) (1-\kappa p_r) (1-\kappa p_\phi)^2 \right]^{-1/2}.
\ee
This can be approximated, up to second order in $\kappa {T^\mu}_\nu$, by
\bq
\tau &=& 1 - \frac12 \kappa (\rho - 3 p) +  \frac{3}{8}\kappa^2 \left[ (\rho - 3 p)^2  +\right.\nonumber  \\
&+& \left. 4 \rho p + 4p_\phi(p_\phi-2p )\right] \,,
\eq 
where we have defined $p = (p_r+2 p_\phi)/3$.

Taking into account that 
\bq
{{\mathcal T}^{0}}_{0} &=& -\frac{1}{{\kappa}}\left(1-\tau\right) - \frac12 \tau (\rho + p_r + 2p_\phi)\,,\\
{{\mathcal T}^{r}}_{r} &=& -\frac{1}{{\kappa}}\left(1-\tau\right) + \frac12 \tau (\rho + p_r-2 p_\phi)\,,\\
{{\mathcal T}^{\phi}}_{\phi} &=& {{\mathcal T}^{\theta}}_{\theta}= -\frac{1}{{\kappa}}\left(1-\tau\right) + \frac12 \tau (\rho - p_r)\,,
\eq
it is simple to calculate the apparent pressure in the perpendicular direction up to first order in $\kappa {T^\mu}_\nu$:
\bq
p_{\perp \rm [A]} &\equiv& \frac13 \left( {{\mathcal T}^{r}}_{r} + {{\mathcal T}^{\phi}}_{\phi}+ {{\mathcal T}^{\theta}}_{\theta}\right)\nonumber\\
&=&  -\frac{1}{{\kappa}}\left(1-\tau\right)+ \frac12 \tau \left(\rho-p \right) \\
&=&p+\kappa p^2 + \frac{\kappa}{8}(\rho+p)^2 + \frac{\kappa}{6}(p_r-p_\phi)^2\,.
\eq

Again, the average value of $p_{\perp \rm [A]}$ inside the particle must be equal to zero, thus implying that the average value of $p$ will be positive or negative, depending, respectively, on whether $\kappa$ is smaller or greater than zero. In fact, one has that
\be 
p=-\frac{\kappa}{8}\rho^2\,,
\ee 
up to first order in $\kappa \rho$. This violation of the von Laue condition for spherically symmetric particles in EiBI gravity was discussed in~\cite{Avelino:2019esh} (which we have followed closely in this section).

As is the case for cosmic strings and domain walls, in EiBI gravity, spherically symmetric particles also have an apparent energy density that differs form the physical one. One may easily check that, up to first order in $\kappa \rho$,
\be 
\rho_{\perp \rm [A]}= \rho -\frac58\kappa\rho^2\,,
\ee 
and so the apparent energy density is smaller or larger than the physical energy density depending on whether $\kappa >0$ or $\kappa <0$.

\section{Conclusions}\label{sec:conc}

In this paper we have studied, both numerically and analytically, domain wall and other defect solutions in the weak-field limit of EiBI gravity. We numerically computed, as a function of $\kappa$, the dependence of the energy density, of the perpendicular pressure, and of the scalar field on the distance to the core of a static planar domain wall, and provided an analytical approximation to the pressure profile valid in this limit. We have also determined analytical approximations to the scalar field profile, and to the physical and apparent energy densities in the vicinity of the domain wall core, valid in the same limit, as well as an estimate of the domain wall tension measured by an outside observer (the so-called apparent domain wall tension). Our results show that EiBI gravity can have a profound impact on the microscopic structure of domain walls, affecting not only their energy density and tension, but also the perpendicular pressure.  As a matter of fact, we have demonstrated that in EiBI gravity the perpendicular pressure may be significantly greater or smaller than zero, depending, respectively, on whether $\kappa$ is positive or negative. The fact that the average perpendicular (physical) pressure is non-zero for $\kappa \neq 0$ constitutes a violation of the generalized von Laue condition, which is in sharp contrast with the $\kappa=0$ case, in which EiBI gravity is indistinguishable from General Relativity and this condition is satisfied.


We have further considered the case of cosmic strings and spherically symmetric particles, showing that EiBI gravity may also have a significant impact on their internal structure. We have found that the breakdown of the generalized von Laue condition at first order in $\kappa \rho$ is also generally expected for cosmic strings and spherically symmetric particles. An exception to this general rule occurs in the case Abelian-Higgs strings with a critical coupling, for which we have found that the von Laue condition holds. Although in this paper we have focused on EiBI gravity, a violation of the generalized von Laue condition should in general be expected in any theory of gravity wherein geometry plays an important role in determining the defect structure. In this sense, EiBI gravity may be regarded as a prototype of gravitational theories in which the strength of gravity can be hugely amplified on microscopic scales with respect to General Relativity. Even though, in the weak-field limit of EiBI gravity a simple reformulation of the von Laue condition in terms of the apparent pressure instead of the physical one would guarantee its preservation, it may be interesting to explore other formulations of the von Laue condition which could be applied in the context of more general theories of gravity.
	
\begin{acknowledgments}
P.P.A. acknowledges the support from Fundação para a Ciência e a Tecnologia (FCT) through the Sabbatical Grant No. SFRH/BSAB/150322/2019. L. S. is supported by FCT  through contract No. DL 57/2016/CP1364/CT0001. Funding of this work has also been provided by FCT through national funds (PTDC/FIS-PAR/31938/2017) and by FEDER—Fundo Europeu de Desenvolvimento Regional through COMPETE2020 - Programa Operacional Competitividade e Internacionaliza{\c c}\~ao (POCI-01-0145-FEDER-031938), and through the research grants UID/FIS/04434/2019, UIDB/04434/2020 and UIDP/04434/2020.
\end{acknowledgments}
 
\bibliography{walls-EiBI}
 	
 \end{document}